\magnification=\magstep1 
\font\bigbfont=cmbx10 scaled\magstep1
\font\bigifont=cmti10 scaled\magstep1
\font\bigrfont=cmr10 scaled\magstep1
\vsize = 23.5 truecm
\hsize = 15.5 truecm
\hoffset = .2truein
\baselineskip = 14 truept
\overfullrule = 0pt
\parskip = 3 truept
\input psfig.sty
\def\frac#1#2{{#1\over#2}}

\nopagenumbers
%This command suppresses the printing of page numbers.
%You should number the pages with blue pencil in upper right corner. 
%
%THE FOLLOWING THREE COMMANDS LEAVE SOME SPACE AT THE TOP OF THE LEAD PAGE. 
%(the command "\vskip 4 truecm" actually results in about 4.5 cm of empty
%space at the top, or about 19.6%).  The publisher will probably reset the 
%chapter heading (your title and by-line), but you should follow my 
%19-20% prescription anyway!  In my design I am following the Les Houches 
%lecture notes volume produced by Nova.   If you have LOTS of authors and 
%by-lines you may want to allow a bit less space at the top (e.g. if you 
%have 3 or more sets of authors and institutions).
%\topinsert
%\endinsert
\vskip 28truecm
\centerline{\bigbfont ABNORMAL OCCUPATION REVISITED}
%If your title is only one line long, put a % before the 2nd title line.
%If your title is longer than two lines, continue thus:
%\vskip 6 truept
%\centerline{\bigbfont AND THE TITLE GOES ON AND ON}
%Don't forget to remove the % sign from the preceding line if you use it! 
\vskip 20 truept
%Now comes your by-line with institutional addresses.  
\centerline{\bigifont J. W. Clark and V. A. Khodel}
\vskip 8 truept
\centerline{\bigrfont McDonnell Center for the Space Sciences} 
\vskip 2 truept
\centerline{\bigrfont and Department of Physics, Washington University}
\vskip 2 truept
\centerline{\bigrfont St. Louis, Missouri 63130, USA} 
%In case of multiple institutions, use the following lines, iterated
%as necessary. 
\vskip 14 truept
\centerline{\bigifont M. V. Zverev}
\vskip 8 truept
\centerline{\bigrfont Russian Research Center Kurchatov Institute}
\vskip 2 truept
\centerline{\bigrfont Moscow 123182, Russia} 
\vskip 1.8 truecm

\centerline{\bf 1.  INTRODUCTION}
\vskip 12 truept

We shall demonstrate that the Fermi surface of dense neutron matter 
may experience a rearrangement near the onset of pion condensation, 
due to strong momentum dependence of the effective interaction induced 
by spin-isospin fluctuations.  In particular, the Fermi surface
may take the form of a partially hollow sphere having a spherical
hole in its center.  Thus, a second (inner) Fermi surface may form 
as high-momentum single-particle states are filled and low-momentum 
states are vacated.  The influence of this phenomenon on the superfluid 
transition temperature of the Fermi system is characterized with the help
of a separation transformation [1,2] of the BCS gap equation.  

The present effort can be viewed as a revival of the search for 
physical realizations of {\it abnormal occupation} in infinite, homogeneous
Fermi systems -- plausible instances in which the quasiparticle distribution 
differs from that of an ideal Fermi gas and Fermi liquid theory breaks 
down.  This search (broadened to Bose systems) was in full swing at 
the time of the Third Workshop on Condensed Matter Theories, held in 
Buenos Aires in 1979, and it was led by Workshop founders Valdir 
Aguilera-Navarro, Ruben Barrera, Manuel de Llano, and Angel Plastino, 
with important contributions from James Vary, John Zabolitzky, and 
others [3-5].  
\vskip 28 truept

\centerline{\bf 2.  THE BUBBLE REARRANGEMENT} 
\vskip 12 truept
Rearrangement of the characteristic Landau quasiparticle distribution
signals a breakdown of Fermi-liquid theory and can lead to profound
alteration of the orthodox behaviors we have come to expect based
on this pervasive physical picture.  For the sake of simplicity, 
our considerations will be limited to homogeneous systems, for which
the Fermi-liquid distribution $n_F(p)$ coincides with the momentum 
distribution of an ideal Fermi gas.  The actual quasiparticle 
distribution $n(p)$ must deviate from $n_F(p)=\theta(p_F-p)$ 
if the necessary condition for its stability is violated.  At
zero temperature, this condition states that the change of the 
ground-state energy $E_0$ remains positive for any admissible 
variation $\delta n(p)$ away from $n_F(p)$.  More specifically,
stability of a given quasiparticle distribution implies
$$
\delta E_0 =\int \xi \left(p,n(p) \right) \delta n(p) 
{{\rm d}^3 p\over (2\pi)^3} > 0 \, , \eqno(1)
$$
where $\xi(p,n(p))\equiv \varepsilon \left(p,n(p) \right)-\mu $ is 
the quasiparticle energy measured with respect to the chemical potential 
$\mu$.  For $n(p)=n_F(p)$, the condition (1) is 
violated if $\epsilon(p)$ rises above $\mu$ at $p<p_F$, or if 
$\epsilon(p)$ falls below $\mu$ at $p >p_F$.  A rearrangement 
of quasiparticle occupancies is precipitated when the density $\rho$ 
attains a critical value $\rho_{cF}$ at which the relation
$$
\xi\left( p,n_F(p);\rho_{cF}\right) =0  \eqno(2)
$$
exhibits a bifurcation leading to a new root $p=p_0$.  This relation 
usually serves only to determine the Fermi momentum $p_F$.  

In homogeneous systems, the simplest type of rearrangement of the momentum 
distribution $n(p)$ of quasiparticles of given spin and isospin
maintains the property that its values are restricted to 0 and 1, but 
the Fermi sea becomes multiply connected (cf.~Refs.\ [4-7]).  
In particular, we may suppose that at densities exceeding the critical 
value $\rho_{cF}$, the normal-state distribution $\theta(p_F-p)$ is
altered by the formation of a ``bubble,'' or particle void,
over a range $p_i < p < p_I < p_F$, with the Fermi momentum $p_F$ 
readjusted to maintain the prescribed  density.  This distribution
is formally represented by 
$n(p)= \theta(p_i-p)+\theta(p_F-p)\theta(p-p_I)$,
where as usual $\theta(x)=1$ if $x\geq 0$ and vanishes otherwise.   
One then has three Fermi surfaces: two inner surfaces 
located at $p_i$ and $p_I$, along with the usual outer surface at $p_F$. 
However, a more dramatic rearrangement can also occur, resulting 
in a distribution with partial occupation of quasiparticle states
that lacks the distinctive trademark of Fermi-liquid theory,
namely a discontinuity of $n(p)$ at the Fermi surface. In
this scenario, called fermion condensation, there exists a finite
range of momenta over which the quasiparticle energy coincides
with the chemical potential, corresponding to the creation of
a ``fermion condensate'' [8-12].

Any change of $n(p)$ from the normal-state distribution $n_F(p)$
must entail an increase of the kinetic energy of the quasiparticle
system.  Accordingly, the anticipated rearrangement only becomes 
possible if it is accompanied by a counterbalancing reduction of 
potential energy, which implies that the effective interaction between 
quasiparticles has acquired a substantial momentum dependence.  The
emergence of such a strong momentum dependence is exactly what one expects 
to occur as the density $\rho$ is increased toward the critical value $\rho_c$ 
for a second-order phase transition in which a branch of the spectrum 
$\omega_s(k)$ of collective excitations of the Fermi system 
collapses at a nonzero value $k_c$ of the wave vector $k$. 
 
To justify this expectation, we follow Dyugaev [13] and consider 
the behavior of the quasiparticle scattering amplitude 
$F({\bf p}_1,{\bf p}_2,{\bf k}) 
\equiv z^2\Gamma({\bf p}_1,{\bf p}_2;{\bf k},\omega=0)M^*/M$ 
near of the phase-transition point.  Here
$\Gamma({\bf p}_1, {\bf p}_2;{\bf k},\omega)$ is the ordinary (in-medium) 
scattering amplitude, $M^*$ is the effective mass, and $z$ is 
the renormalization factor specifying the weight of the quasiparticle
pole.  The amplitude $F$ can be written as the sum $F^r+F^s$ of a 
regular part $F^r$ and a singular part $F^s$, with the latter
taking the universal form 
$$
F^s_{\alpha\delta;\beta\gamma}({\bf p}_1,{\bf p}_2,{\bf k};
\rho\rightarrow \rho_c)= -O_{\alpha\delta}O_{\beta\gamma} D(k)+
O_{\alpha\gamma}O_{\beta\delta}D(|{\bf p}_1-{\bf p}_2+{\bf k}|)  
\eqno(3)
$$
in terms of the propagator $D(k)$ of the collective excitation. 
This form has been derived with due attention to the antisymmetry
of the two-particle wave function under exchange of the particle 
coordinates (spatial, spin, isospin).  The collective propagator 
is conveniently parametrized according [13]
$$
D^{-1}(k)=\beta^2+\gamma^2(k^2/k^2_c-1)^2 \, ,
\eqno(4)
$$
where the parameter $\beta(\rho)$, with $\beta(\rho_c)=0$,
measures the proximity to the phase-transition point.  The vertex
$O$ appearing in Eq.~(3) determines the structure of the 
collective-mode operator and is normalized by ${\rm Tr}(OO^\dagger)=1$. 
Specifically, the choice $O=1$ is made in treating the rearrangement 
of the quasiparticle distribution due to collapse of density 
oscillations [14], while $O={\vec \sigma}$ is appropriate 
when studying the rearrangement of $n_F(p)$ triggered by the 
softening of the spin collective mode [15].  In the present 
investigation we will be concerned with dense, homogeneous neutron 
matter in which abnormal occupation is induced by spin-isospin 
fluctuations; thus the pertinent operator is 
$O= ({\vec \sigma} \cdot {\bf k}){\vec \tau}$.

Cutting through the details, the most important features of the 
model defined by Eqs.~(3) and (4) are that the function 
$F^s({\bf p}_1,{\bf p}_2, {\bf k}=0)\simeq D({\bf p}_1-{\bf p}_2)$   
depends on the difference ${\bf p}_1-{\bf p}_2$ and that in the 
neighborhood of the soft-mode phase-transition point this dependence 
becomes very strong. 

Eqs.~(3) and (4) furnish a suitable basis for efficient evaluation of 
the single-particle spectrum $\xi(p)$ near the second-order phase 
transition.  We exploit a straightforward connection between $\xi(p)$ 
and the scattering amplitude $F({\bf p}_1,{\bf p}_2,{\bf k} =0)$, 
thereby circumventing the awkward frequency integration that would 
be encountered in an RPA approach.  This connection is made 
through the relation
$$
{\partial \xi(p)\over \partial {\bf p}}={{\bf p}\over M}
+{1\over 2}\int F_{\alpha\beta;\alpha\beta} ({\bf p},{\bf p}_1) 
{\partial n(p_1)\over \partial {\bf p}_1} {{\rm d}^3p_1\over(2\pi)^3} 
\, ,
\eqno(5)
$$
which may be derived by means of the Landau-Pitaevskii identities 
[16-18].  The contribution to Eq.~(5) from the singular 
part (3) of $F$ can be easily integrated over the momentum 
$\bf p$ to obtain
$$
\xi(p)= {p^2\over 2M_r^*}+{1\over 2} \int
D({\bf p}-{\bf p}_1) n(p_1) {{\rm d}^3 p_1\over (2\pi)^3} 
\, .
\eqno(6)
$$
In stating this result, we assume that the contribution to 
$\xi(p)$ from the {\it regular, nonsingular} part 
of $F$ can be simulated by replacing of the bare mass $M$ 
appearing in Eq.~(5) by an effective mass $M^*_r$.  The generally 
accepted values for this effective mass are in the range 0.7--0.8 
for the pertinent densities in the neutron-star interior.

According to Migdal and collaborators [19,20] (see also 
Ref.~[21]), a dramatic phase transition can occur when the density
$\rho$ of neutron matter in the liquid core of a neutron star reaches
a critical density $\rho_{c\pi}$ of some 2--3 times the equilibrium
density $\rho_0$ of symmetrical nuclear matter.  The spin-isospin collective 
mode collapses at a finite wave vector $k=k_c\sim p_F$ and a phase 
transition identified as pion condensation sets in.  A prominent
feature of the ground state of the system beyond the phase-transition 
point is the presence of a condensate of spin-isospin density waves.  
It should be clear from Eqs.~(3) and (4) that spin-isospin 
fluctuations with $k\sim k_c$ are strongly amplified in the 
neighborhood of the transition as a consequence of the divergence 
of the propagator $D(k\to k_c,\rho_c)$. 

We are thus led to apply Eq.~(6) to dense neutron matter 
close to the onset of neutral pion condensation.  Employing the
parametrization (4) in Eq.~(6), we obtain the working formula
$$
\xi(p)= {p^2\over 2M^*_r}+ {1 \over 2}\int 
{1\over \beta^2 +\gamma^2(({\bf p}-{\bf p}_1)^2 -k_c^2)^2/ k_c^4}n(p_1) 
{{\rm d}^3p_1\over (2\pi)^3} \,.
\eqno(7)
$$
Ideally, one would like to extract quantitatively reliable values
for the input parameters $\beta$, $\gamma$, and $k_c$ from a
convincing {\it ab initio} treatment of neutron-star matter.
Unfortunately, no such treatment is yet available.  Moreover, 
current predictions of the critical density $\rho_{c\pi}$
range from 0.2 to 0.5 fm$^{-3}$ (i.e., some 1--3 times $\rho_0$),
depending on what theoretical assumptions are implemented [19-21].

This situation leaves us with no alternative but to carry out
calculations for several representative choices the parameters of 
the microscopic model.  Inserting the formula (7) into 
Eq.~(2), we determine the critical density $\rho_{cF}$ at which 
the solution of Eq.~(2) bifurcates.  For $\rho > \rho_{cF}$ 
this equation then yields two new momenta $p_i$ and $p_I$ at which
$\xi(p)$ vanishes, and between which $\xi(p)$ is positive.  The
bubble region lies between these two momenta.  

The necessity for brevity in this presentation precludes the explicit 
presentation of the numerical results that have been derived for the 
spectrum $\xi(p)$ and for the phase diagram of dense neutron matter. 
The neutron spectrum has been calculated at the critical densities 
$\rho_{cF}$ corresponding to three different 
sets of model parameters: (a) $\gamma=1.25 m_\pi$, $k_c=0.9\,p_F$, 
$\beta^2=0.22\,m_{\pi}^2$ ($\rho_{cF}\simeq 1.19\,\rho_0)$, 
(b) $\gamma=1.25 m_\pi$, $k_c=0.9\,p_F$, $\beta^2=0.25\,m_{\pi}^2$ 
($\rho_{cF}\simeq 1.76\,\rho_0)$, and (c) $\gamma=1.25 m_\pi$, $k_c=p_F$, 
$\beta^2=0.13\,m_{\pi}^2$ ($\rho_{cF}\simeq 1.88\,\rho_0)$, where
$m_\pi$ is the pion mass.  The choice $k_c= 0.9p_F$ for the critical 
wave number is suggested by earlier numerical investigations [20].  
Two different positions were found for the bifurcation point, namely 
$p_0=0$ (for parameter sets (a) and (b)) and $p_0\simeq 0.12\,p_F$
(for set (c)).  
The phase diagram of neutron matter in the variables $\rho$ (measured 
in $\rho_0$) and $\beta^2$ (measured in $m_{\pi}^2$) has been 
constructed at $k_c=0.9\,p_F$ for four different values of $\gamma$ 
(1.0, 1.2, 1.4, and 1.6, in units of the pion mass).  Plots 
of the results of these calculations may be found in Refs.~[18,22].

Variation of the parameters $\beta$, $\gamma$,
and $k_c$ within sensible bounds can significantly affect the phase
diagram and hence the extent, in density, of the phase with rearranged
quasiparticle occupation.  Even so, our numerical study has revealed
four characteristic and generic features of the bubble rearrangement. 
\item{(i)} The critical density $\rho_{cF}$ for the rearrangement is 
less than the critical density $\rho_{c\pi}$ for pion condensation.  
Since both phenomena arise from the strong momentum dependence of the 
amplitude $F({\bf p}_1,{\bf p}_2;{\bf k}\rightarrow 0)$,
rearrangement of the quasiparticle distribution may be regarded as
a {\it precursor} of pion condensation.  
\item{(ii)} The bifurcation point associated with formation of a 
bubble in the neutron momentum distribution is located at 
small momenta, $p_0< 0.2p_F$, regardless of the applicable value 
of $\rho_{c\pi}$.  
\item{(iii)} The spectrum $\xi(p)$ exhibits a deep depression for 
$p \sim (0.5-0.6)p_F$.  
\item{(iv)}
The ratios $\rho_{cF}/\rho_{c\pi}$ and $p_0/p_F$ are insensitive 
to the actual value taken by $\rho_{c\pi}$ within the range of 
plausible theoretical predictions. 

The emergence in neutron matter of one or more new Fermi surfaces 
positioned at low momentum values would provide a new avenue for
rapid direct-Urca neutrino cooling of neutron stars [22].
More broadly, the creation of new Fermi surfaces by the mechanism
we have described -- as well as the more profound rearrangement
involved in fermion condensation -- would call for revision
of many of the conclusions that have been developed within
Fermi liquid theory.  Here we shall focus on some elementary
properties of pairing in the reconfigured system.
\vskip 28 truept

\centerline{\bf 3.  PAIRING IN THE PRESENCE OF ABNORMAL OCCUPATION}
\vskip 12 truept

We assume, for the sake of simplicity, that beyond the instability point 
there exist only two Fermi surfaces, an outer one corresponding to the usual 
Fermi momentum $p_F$ and an additional inner one at $p_I$, lying 
close to the origin in momentum space.  Hence we consider the
limiting case $p_i=0$ in our original specification of the ``bubble'' 
rearrangement.  Also for the sake of simplicity, we restrict 
the analysis to $^1$S$_0$ pairing, for which the BCS gap equation 
has the familiar form
$$
\Delta(p)=-\int {\cal V}( p, p_1){\cal E}^{-1}(p;T)\Delta(p_1)
{\rm d}\tau_1
\, , \eqno(8)
$$
where ${\cal V}$ is the effective particle-particle interaction
and we employ notations $d \tau = p^2 dp /2 \pi^2$ for 
the volume element and 
${\cal E}^{-1}(p;T)=  \left[2E(p)\right]^{-1}\tanh \left[ E(p)/ 2 T \right]$ 
for the usual combination of $\tanh$ temperature factor and energy 
denominator $2E(p)$.  The appearance of the gap function $\Delta(p)$ 
in the superfluid quasiparticle energy 
$E(p)=\left[\xi^2(p)+\Delta^2(p)\right]^{1/2}$ renders the
gap problem nonlinear.  The quantity $\xi(p)$ is 
to be interpreted as the single-particle spectrum in the system with 
pairing turned off. 

Adopting the strategy for solving gap equations that was introduced 
in Ref.~[1] and elaborated in Refs.~[2,3], we write
the block ${\cal V}$, identically, as a separable part plus a remainder 
that automatically vanishes on the outer Fermi surface.
Hence we write
$$
{\cal V}(p_1,p_2)\equiv V_F\phi_F(p_1)\phi_F(p_2)+W(p_1,p_2)
\eqno(9)
$$
and take $\phi_F(p)= {\cal V}(p,p_F)/ V_F$, where $V_F\equiv 
{\cal V}(p_F,p_F)$.  It follows directly that 
$W(p,p_F)\equiv W(p_F,p)= 0$, as required.  In the ordinary case
where there is only one Fermi surface, this decomposition allows 
us to replace the singular nonlinear integral equation (8) 
by two equivalent equations: (i) a nonsingular quasilinear 
integral equation for a $T$-independent shape factor $\chi(p)
= \Delta(p)/\Delta_F$ and (ii) a nonlinear `algebraic' equation for the
$T$-dependent gap value $\Delta_F(T)\equiv \Delta(p_F,T)$.
In the present case where there are two Fermi surfaces, we must 
extend the procedure of Ref.~[1] to deal consistently with
the inner Fermi surface as well as the outer one.  This is
accomplished by decomposing the interaction term $W$ appearing 
in Eq.~(9) in the same manner as before, setting
$$
W(p_1,p_2)\equiv W_I\phi_I(p_1)\phi_I(p_2)+Y(p_1,p_2) 
\eqno(10)
$$
with $\phi_I(p)=W(p,p_I)/W_I$ and $W_I=W(p_I,p_I)\equiv 
{\cal V}(p_I,p_I)-{\cal V}^2(p_F,p_I)/{\cal V}(p_F,p_F)$, so 
that $Y(p,p_I)\equiv Y(p_I,p)\equiv Y(p_F,p)\equiv Y(p,p_F)= 0$. 
The above relations entail the boundary values
$$
\phi_F(p_F)=1, \qquad \phi_I(p_I)=1, \qquad \phi_I(p_F)=0 \, ,
\eqno(11)
$$
while the key quantity $\phi_F(p_I)\propto {\cal V}(p_I,p_F)$ 
describes the connection between the quasiparticles of the two
Fermi surfaces in the particle-particle channel.  If $\phi_F(p_I)$ 
vanishes, these surfaces are disconnected and the problem is 
trivialized.

In the general case where ${\cal V}(p_I,p_F)\neq 0$, substitution
of Eqs.~(9) and (10) into the BCS gap equation (8) gives
$$
\eqalignno{
\Delta(p)&=- V_F\phi_F(p)\int \phi_F(p_1){\cal E}^{-1}(p_1;T) 
\Delta(p_1) {\rm d} \tau_1  -W_I\phi_I(p)\int \phi_I(p_1) 
{\cal E}^{-1}(p_1;T) \Delta(p_1) {\rm d} \tau_1 \cr
& \qquad - \int Y( p, p_1){\cal E}^{-1}(p_1;T)\Delta(p_1) {\rm d} \tau_1 \, .
&(12) \cr
}
$$
This equation is conveniently rewritten as
$$
\Delta(p)=B_F\chi_F(p) +B_I\chi_I(p) \, ,
\eqno(13)
$$
with
$$
\eqalignno{
B_F&=- V_F\int \phi_F(p){\cal E}^{-1}(p;T)\Delta(p){\rm d}\tau \,, \cr
B_I&=-W_I\int \phi_I(p) {\cal E}^{-1}(p;T) \Delta(p) {\rm d} \tau \, ,
&(14) \cr }
$$
and
$$
\eqalignno{
\chi_F(p)&=\phi_F(p)-\int Y( p, p_1) {\cal E}^{-1}(p_1;T)
\chi_F(p_1) {\rm d} \tau_1 \, , \cr
\chi_I(p)&=\phi_I(p)-\int Y( p, p_1) {\cal E}^{-1}(p_1;T) 
\chi_I(p_1) {\rm d} \tau_1 \, .
&(15) \cr}
$$
Referring to the relations (11), we observe that 
$$
\chi_I(p_I)=\chi_F(p_F)=1\,, \qquad \chi_I(p_F) =0\,, 
\qquad \chi_F(p_I)=\phi_F(p_I)= {\cal V}(p_I,p_F)/{\cal V}(p_F,p_F)\,, 
\eqno(16)
$$
because the block $Y$ is zero when either of its arguments lies on 
a Fermi surface.  By this same property, it is permissible, 
inside the quantity ${\cal E}^{-1}$ appearing in the integral 
equations (15), to replace the superfluid quasiparticle 
energy $E(p_1)$ by $|\xi(p_1)|$ and the temperature factor 
$\tanh\left[ E(p_1) /2T \right]$ by unity.  Because the energy
gaps involved are generally quite tiny compared to the Fermi energy,
these replacements are valid to a superb approximation.  We are
left with the linear integral equations
$$
\eqalignno{
\chi_F(p)&=\phi_F(p)-\int Y( p, p_1){1\over 2|\xi(p_1)|}
\chi_F(p_1) {\rm d} \tau_1 \, , \cr 
\chi_I(p)&=\phi_I(p)-\int Y( p, p_1){1\over 2|\xi(p_1)|}\chi_I(p_1)
{\rm d} \tau_1 \, , & (17) \cr
}
$$
for the two shape functions needed to construct the gap function
$\Delta(p)$ using Eq.~(13).  Since there remains no trace of 
the temperature $T$ in Eqs.~(17), we are free to regard the 
solutions $\chi_I(p)$ and $\chi_F(p)$ as $T$-independent quantities.

Appealing to the properties (16), Eq.~(13) yields
$$
\Delta_F\equiv\Delta(p_F)=B_F\,, 
\qquad \Delta_I\equiv\Delta(p_I)=B_I+B_F\phi_F(p_I) \, .
\eqno(18)
$$
Inserting the decomposition (13) into Eqs.~(14), 
we arrive at a system of two equations
$$
\eqalignno{
B_F&=- V_F L_{FF}B_F-V_F L_{FI}B_I \, , \cr
B_I&=-W_I L_{IF}B_F-W_I L_{II}B_I \, ,
&(19) \cr }
$$
for determination of the amplitudes $B_F$ and $B_I$ entering
Eq.~(13), where
$$
L_{FF}=\int \phi_F(p) {\cal E}^{-1}(p;T) \chi_F(p) {\rm d} \tau \,, \quad 
 L_{II}=\int \phi_I(p) {\cal E}^{-1}(p;T) \chi_I(p) {\rm d} \tau  \, , 
$$
$$
L_{IF}\equiv L_{FI}=\int \phi_I(p){\cal E}^{-1}(p;T) \chi_F(p) {\rm d} \tau 
\equiv \int \phi_F(p) {\cal E}^{-1}(p;T) \chi_I(p) {\rm d} \tau \, .
\eqno(20)
$$
It is helpful to recast the system (19) in an equivalent form
$$
\left[1+V_F L_{FF}(T)-V_F\phi_F(p_I) L_{FI}(T)\right]\Delta_F+
 V_FL_{FI}(T)\Delta_I \, ,
$$
$$
\left[ W_I L_{FI}(T)-(1+W_I L_{II}(T))\phi_F(p_I)\right]\Delta_F
+[1+W_I L_{II}(T)]\Delta_I = 0 \,  .
\eqno(21)
$$
For a solution to exist, the determinant ${\cal D}(T)$ of (19) or 
(21) must equal zero for any $T$.  
Together with either of the two equations (19) [or either of 
(21)], the dispersion relation ${\cal D}(T)=0$ forms a 
closed system that permits one to determine all characteristics of 
the superfluid system feeding upon the two Fermi surfaces located at 
$p_I$ and $p_F$.  

We begin to explore the implications of the formalism we have 
developed by examining the effect of the additional (inner) 
Fermi surface on the superfluid transition temperature $T_c$. 
Observe that Eqs.~(19) [or (21)] become decoupled if $L_{IF}=0$.  
Assume, as a first case (Case I), that {\it both} of the interaction
parameters $V_F$ and $W_I$ are negative, so that Cooper pairing could 
exist at {\it both} Fermi surfaces when they are disconnected.  The 
pairing effect is naturally stronger at the ``main'' or outer surface due 
to a greater density of states.  From the two solutions of the problem 
as stated, we therefore select $\Delta(p)=\Delta_F \chi_F(p)$ with 
$\Delta_I=\phi_F(p_I)\Delta_F$, implying that the individual critical 
temperatures $T^F_c$ and $T_c^I$ satisfy $T^F_c>T^I_c$.  It should
be noted that in spite of this inequality, the magnitude of the ratio 
$\Delta_I/\Delta_F\sim \phi_F(p_I)$ is not necessarily less than 
unity (see below).

Working in the vicinity of the transition temperature $T_c$, standard 
arguments and manipulations in the spirit of BCS theory reveal the
behaviors 
$$
\eqalignno{
L_{FF}(T\to T_c)&\rightarrow N_F(0)\left\{(1+ g^2_{IF})(L+ \alpha_1\tau)  
-\alpha_2 \left[ D^2_F+g^2_{IF}D^2_I\right]\right\} \, , \cr
L_{II}(T\to T_c)&\rightarrow N_I(0)\left[L+ \alpha_1\tau -\alpha_2 
D^2_I\right] \, , \cr
L_{IF}(T\to T_c)&\rightarrow N_I(0)\phi_F(p_I)\left[L+\alpha_1\tau-\alpha_2 
D^2_I\right]\, , &(22) \cr }
$$
in terms of the three dimensionless parameters $\tau=(T_c-T)/T_c$, 
$D_F=\Delta_F/T_c$, and $D_I=\Delta_I/T_c$.  In Eqs.~(22), $N_F(0)$ 
and $N_I(0)$ are respectively the densities of states at the outer 
and inner Fermi surfaces, $g^2_{IF}=\phi_F^2(p_I)N_I(0)/N_F(0)$ 
is an effective coupling constant, and 
$L=\ln(\varepsilon^0_F/\pi T_c)+ C$ measures the transition
temperature, where $\varepsilon_F^0$ is the free Fermi energy and
C is Euler's constant (0.577).  Certain constants entering the 
derivation of the limiting behaviors of $L_{II}(T)$, $L_{IF}(T)$, 
and $L_{FF}(T)$ have no material role in our arguments; they
merely effect a renormalization of the critical temperatures 
$T^F_c$ and $T^I_c$ and may be thus be suppressed in forming 
Eqs.~(22).  The relevant temperature dependences are determined 
entirely by the ratio $\alpha_1/\alpha_2=8\pi^2/7\zeta(3)$.

After substituting the results (22) into Eqs.~(21), we 
arrive at
$$
\left[1+V_FN_F(0)(L+\alpha_1\tau-\alpha_2 D^2_F)\right]D_F+V_FN_I(0)\phi_F(p_I)
(L+\alpha_1\tau-\alpha_2 D^2_I)D_I=0 \, ,
$$
$$
-\phi_F(p_I)D_F+\left[1+W_IN_I(0)\right](L+\alpha_1\tau-\alpha_2 D^2_I) D_I=0
\,.
\eqno(23)
$$
Setting $T=T_c$ and evaluating the determinant ${\cal D}(T_c)$ of
this system, we obtain a closed formula for the new critical 
temperature $T_c$ in terms of the individual critical temperatures
$T_c^F$ and $T_c^I$ for the uncoupled system:
$$
(L-l_I)(L-l_F)-g_{IF}^2l_IL = 0 \, .
\eqno(24)
$$
The constants $l_F$ and $l_I$ entering this relation are defined by
$l_F\equiv \ln(\varepsilon_F^0/\pi T_c^F)+C=-1/V_FN_F(0)$ and
$l_I\equiv \ln(\varepsilon^0_F/\pi T_c^I)+C=-1/W_IN_I(0)$. 
The inequality $T_c^F > T_c^I$ clearly implies $l_I>l_F$.

The situation at small coupling, $g^2_{IF}<<1$, is especially transparent.   
In this case the quantity $L$ (which measures $T_c$) is not much
different from $l_F$ (which measures $T_c^F$), allowing us to replace $L$ 
by $l_F$ in the last term of the determinantal condition (24).  
The solution of Eq.~(24) is then 
$$
L_{\pm}={l_I+l_F\over 2} \pm \left[{(l_I-l_F)^2\over 4}+
g_{IF}^2l_Il_F\right]^{1\over 2} \, .
\eqno(25)
$$
This result reminds us of the familiar textbook solution of the 
two-level problem.  In that problem, the two energy levels repel each other
when the off-diagonal interaction is switched on.  The lower
level moves downward and the upper level moves upward.  In 
the current problem, the greater logarithm (in this case, $L_+$) 
increases while the smaller logarithm ($L_-$) decreases.  In 
particular,
$$
L_- -l_F\simeq -g^2_{IF}{l_Il_F\over l_I-l_F} \, .
\eqno(26)
$$
In the case being considered, $l_I$ and $l_F$ are both positive, 
with $l_I>l_F$.  We may therefore conclude that the emergence of 
the second Fermi surface {\it increases} the critical temperature $T_c$ 
relative to $T_c^F$. 

The picture changes nontrivially when we consider the more interesting 
case (Case II) in which pairing is {\it absent} at the new Fermi surface 
when the two surfaces are disconnected, but is still present at the 
original surface.  When the coupling is turned on, a pairing gap is 
found to exist on the {\it new} Fermi surface as well as the old one, a
feature that is directly evident from either of the equations (21).  
The result (25) continues to apply (again assuming small coupling).  
However, in contrast to Case I, the value of $l_I$ becomes negative 
while $l_F$ remains positive.  Consequently, the single acceptable
value of $L$ derived from Eq.~(25) increases relative
to $l_F$, implying a {\it decrease} of $T_c$ with respect to $T_c^F$.  
This behavior should not be surprising: the value of the pairing 
gap depends on the shape of the single-particle spectrum, and if the 
spectrum becomes flatter in a region where the interaction is repulsive, 
there must be a suppression of the gap value and a concomitant
suppression of $T_c$.  We must stress that the situation is now 
quite different from that of perturbation theory, where the gap 
increases independently of the sign of the perturbating interaction.  
The distinctive behavior we have described is indicative of a failure 
of perturbation theory in Case II.  We should also point out 
the close resemblance between the predicted behavior and the proximity 
effect observed in junctions between a superconductor and a normal 
metal: the superconductor tends to induce superconditivity on the 
normal side of the junction, while suppressing its strength on 
the superconducting side.

Here we have confined our attention to the effect of the bubble
rearrangement on the superfluid transition temperature $T_c$, considering
the possible scenarios for pairing when there are two concentric 
Fermi surfaces.  Results are also available [18] for the modifications
produced in the specific-heat discontinuity at $T_c$ and for
the relation between $T_c$ and the energy gap $\Delta_F$ at $T=0$.
\vskip 28truept

\centerline{\bf 4.  CONCLUSIONS}
\vskip 12truept

We have studied the rearrangement of single-particle degrees 
of freedom that precedes the onset of pion condensation,
and we have found that this rearrangement may express
itself in the emergence of a bubble in the quasiparticle momentum 
distribution, i.e., the formation of a new Fermi sea with a
spherical hole in the middle.  This is in fact one of the scenarios
considered some two decades ago by de Llano, Plastino, and their
collaborators.  The formalism we have developed 
and the results we have obtained can be applied more widely in 
the theory of strongly correlated Fermi systems.  In this spirit, 
it will be of special interest to re-examine the case of superfluid $^3$He, 
which offers a realistic example of a Fermi liquid existing near an 
antiferromagnetic phase transition.  In our view, rearrangements
of the single-particle degrees of freedom arising from momentum-dependent
effective interactions have a generic character.  Such rearrangements 
include not only bubble configurations, but also the phenomenon of 
fermion condensation [8-12].  We call attention especially 
to Ref.~[11], where the competition between bubble rearrangement and 
fermion condensation is studied under variation of the temperature.
\vskip 28 truept

\centerline{\bf ACKNOWLEDGMENTS}
\vskip 12 truept

This paper is dedicated, in admiration and friendship, to
Manuel de Llano and Angel Plastino as they reach the venerable
age of sixty.  The research was supported in part by the 
U.S.\ National Science Foundation under Grant No.~PHY-9900713 (JWC 
and VAK), by the McDonnell Center for the Space Sciences (VAK), and 
by Grant No.~00-15-96590 from the Russian Foundation for Basic 
Research (VAK and MVZ).  MVZ is grateful for hospitality extended by 
the INFN (Sezione di Catania).  JWC thanks the US Army Research 
Office for travel support through a grant to Southern Illinois 
University--Carbondale.
\vskip 28 truept

\centerline{\bf REFERENCES}
\vskip 12 truept
\item{[1]}
V.~V.~Khodel, V.~A.~Khodel, and J.~W.~Clark, {\it Nucl.~Phys.} {\bf A598},
390 (1996).
\item{[2]}
V.~V.~Khodel, V.~A.~Khodel, and J.~W.~Clark, {\it Phys.~Rev.~Lett.}
{\bf 81}, 3828 (1998).
\item{[3]}
V.~A.~Khodel, V.~V.~Khodel, and J.~W.~Clark, {\it Nuc.~Phys.~A}, in press.
\item{[4]} 
M.~de Llano, {\it Nucl.~Phys.} {\bf A317}, 183 (1979).  
\item{[5]} 
M.~de Llano and J.~P.~Vary, {\it Phys.\ Rev.\ C} {\bf 19}, 1083 
\item{[6]} V. C. Aguilera-Navarro, M. de Llano, J. W. Clark, and
A. Plastino, {\it Phys.\ Rev.\ C} {\bf 25}, 560 (1982).
\item{[7]} M. de Llano, A. Plastino, and J. G. Zabolitzky, 
{\it Phys.\ Rev.\ C} {\bf 20}, 2418 (1979); {\bf 22}, 314 (1980). 
\item{[8]} V.~A.~Khodel and V.~R.~Shaginyan, {\it JETP Lett.} {\bf 51}, 553 
(1990); {\it Condensed Matter Theories} {\bf 12}, 222 (1997).
\item{[9]} P.~Nozieres, {\it J.~Phys.~I France} {\bf 2}, 443 (1992).
\item{[10]}V.~A.~Khodel, J.~W.~Clark, and V.~R.~Shaginyan, {\it Solid
State Comm.} {\bf 96}, 353 (1995).
\item{[11]} M. V. Zverev and M. Baldo, {\it JETP} {\bf 87}, 1129 (1998).
\item{[12]}M. V.~Zverev, V.~A.~Khodel, and M.~Baldo, {\it JETP Lett.} {\bf 72},
126 (2000).
\item{[13]} A.~M.~Dyugaev, {\it Sov.\ Phys.} {\it JETP} {\bf 43}, 1247 (1976).
\item{[14]} V.~A.~Khodel, V.~R.~Shaginyan, and M. V.~Zverev, {\it JETP Lett.}
65, 253 (1997)
\item{[15]}M. V.~Zverev and V.~A.~Khodel, 
http://xxx.lanl.gov/cond-mat/9907061.
\item{[16]} L.~P.~Pitaevskii, Sov.\ Phys.\ {\it JETP} {\bf 10}, 1267 (1960).
\item{[17]} A.~A.~Abrikosov, L.~P.~Gor'kov and I.~E.~Dzialoshinskii, 
{\it Methods of Quantum Field Theory in Statistical Physics}
(Prentice-Hall, Englewood Cliffs, NJ, 1965). 
\item{[18]} J.~W.~Clark, V.~A. Khodel, and M. V. Zverev, {\it Sov. J. Nucl.
Phys.}, in press. 
\item{[19]} A.~B.~Migdal, {\it Rev.~Mod.~Phys.} {\bf 50}, 107 (1978).
\item{[20]} A.~B.~Migdal, E.~E.~Saperstein, M.~A.~Troitsky, and 
D.~N.~Voskresensky, {\it Phys.~Rep.} {\bf 192}, 179 (1990).
\item{[21]}
A.~Akmal, V.~R.~Pandharipande, D.~G.~Ravenhall, {\it Phys.~Rev.\ C} {\bf 58}, 
1804 (1998). 
\item{[22]} D.~N.~Voskresensky, V.~A.~Khodel, M.~V.~Zverev, and 
J.~W. Clark, {\it Ap.~J.} {\bf 533}, L127 (2000).
\end